# Martian gullies: Produced by fluidization of dry material

Yolanda Cedillo-Flores, [a,b] and [a]Héctor Javier Durand-Manterola

yolanda@geofisica.unam.mx , hdurand_manterola@yahoo.com
[a] Departamento de Física Espacial, Instituto de Geofísica, Universidad Nacional Autónoma de México. UNAM. Coyoacan C.P 04510. D.F. Mexico.
[b] Posgrado de Geografía, Facultad de Filosofía y Letras, Universidad Nacional Autónoma de México. Circuito Interior. Ciudad Universitaria s/n. C.P 04510. D.F. Mexico.

Abstract
*Context:* The gullies on Mars were discovered in the year 1999. Their aspect suggests that they had been formed recently on Martian slopes.
Since then several hypotheses have appeared trying to explain the presence of these gullies. The main hypotheses are the ones which suggest that some liquid, water or $CO_2$, was responsible for modeling the gullies and ones that propose dry flows as the modeling agents.
*Objective:* The aim of this work is to develop an alternative hypothetical mechanism of formation of Martian gullies.
*Method:* Our model proposes that the Martian gullies were formed as a result of a fluidization process of the material deposited on the slopes of the impact craters, plateaux and other geomorphologic structures. This fluidization is caused by the sublimation of carbon dioxide ice deposited in the form of snow, due to the daily and seasonal temperature changes. We also present the results of an experimental simulation.
*Results:* Structures similar to the Martian gullies were reproduced using the air injection mechanism, as a substitute to gaseous $CO_2$, on a sandy slope. The Reynolds number for our experimental flow and for the flow in the Martian gullies was calculated and we found that they are of the same order, whilst the water flows have much higher Reynolds numbers.

Taking into account the current environmental conditions for Mars, from our results it may be suggested that this mechanism is the possible modeling agent for the Martian gullies. Two of the most important characteristics of the proposed model are: a) It offers a simple explanation of how recurrent fluidization events occur in the same place, a necessary recurrence of the formation of a gully, offering a recharging mechanism for the system after every fluidization event, and b) It considers the formation of gullies possible even in angles smaller than the angle of repose. These characteristics of the model solve these two problems present in current theories.

## 1 Introduction:

In the year 1999, Malin and Edgett discovered gullies in the images from planet Mars, obtained by the Mars Global Surveyor (MGS) spacecraft, which had the appearance of being formed recently on Martian slopes. (Malin and Edgett, 2000; Hartmann, 2003).

The discovery of Martian gullies immediately generated the following questions: How did these gullies form? What structure did they have? How old were they? On what type of geological material were they formed? What was the geographic distribution of these features?

Gullies are landforms formed over several types of slopes. The Martian gullies normally exhibit the three characteristic forms present in terrestrial gullies: alcove, channel and apron or debris cone, although not all the gullies present these three forms or features. The channels extend from the base of the alcove towards the lower part of the slope. Some of the channels gradually fade at the foot of the slope.

These features are located in the relief of the Noachian, Hesperian and Amazonian eras in both hemispheres, north and south, as well as in several latitudes. However the gullies predominate in mid-latitudes, between the 30° and 60° (Heldmann and Mellon, 2004, Mellon and Phillips, 2001; Bridges and Laqckner, 2006). Checking images obtained from the cameras of three NASA spacecrafts: Martian Orbiter Camera (MOC), Mars Global Surveyor (MGS), Thermal Emission Imaging System (THEMIS), from the Mars Odissey spacecraft, and High Resolution Imaging Science Experiment (HiRISE), from the Reconnaissance Orbiter spacecraft, as well as some images obtained by the Mars Express spacecraft, from the ESA, (European Spatial Agency) we led to the same conclusions.

Several explanations of the origin of the gullies have emerged which can be separated into two groups: theories which invoke the presence of a fluid (generally water) and theories which proposes the existence of the gullies as a result of dry material avalanches.

The first theories include the existence of aquifers, (Mellon and Phillips, 2001; Frey et al, 2004; Costard et al, 2002; Christensen et al., 2003; Mangold et al., 2003), highly saline water (Doran y Forman, 2000; Knauth et al., 2000; Wynn-Williams et al., 2001; Knauth and Burt, 2002; Heldmann et al., 2005) or liquid carbon dioxide (Musselwhite et al, 2001; Draper et al., 2000 y Hoffman, 2001 a and b). The second group of theories refers to avalanches of deposited granular material by means of aeolian activity (Treiman 2003, Shinbrot, et al,. 2004).

The argument for the gullies having been formed by water or liquid $CO_2$ presents several weak points due to the pressure and temperature conditions which currently exist on Mars and the distribution of these features. In the distant past, when the atmosphere and pressure was high enough to maintain the water in a liquid state on the surface of the planet, the formation of the gullies by this liquid was feasible. However, characteristics of the gullies indicate that they are probably not that old, maybe some millions of years old. The prevailing conditions on Mars in recent millions of years make it very improbable that water is the main agent responsible for the formation of gullies. Although Lobitz et al (2001) have demonstrated that liquid water can be stable over extended areas on the surface of Mars, the excavation of gullies on rocky slopes requires several water liberation events from an aquifer and this in turn requires its own refilling mechanism. On Earth this refilling is done by rain, but this option is beyond the possibilities of the current Martian environment.

On the other hand, the dry flow theory also presents weak points. Most of the gullies have been formed on slopes with inclinations smaller than the angle of repose (Heldmann y Mellon, 2004) and it is not quite clear how the dry material could have slid down under these conditions, which represents an unresolved point for the dry flow theory.

Pelletier et al (2008) performed simulations and compared velocity with distance between the flow with liquid and the dry flow. The end result was that the dry flow can erode to angles smaller than 33°, whilst it preserves its momentum, but it is necessary that it commences at angles greater than 33°.

In this work an alternative model is proposed to try to solve the weak points present in other hypotheses. We suggest that the Martian gullies were formed, and still are formed, by fluidization of dust and water ice accumulated on the impact crater slopes or canyons due to the sublimation of $CO_2$ snow presents on these slopes as well.

A series of experimental simulations were performed using an experimental device in order to try to reproduce analogue forms of Martian gullies. Then the results were compared.

## 2 Model and experimental simulation
### 2.1 Gullies formed by granular flows fluidized by $CO_2$ gas.

In the proposed model, the scenario is as follows: over the gullies of the impact craters, canyons and other Martian landforms, the wind deposits dust and sand. This material then slides down towards the base of the slope, the resulting erosion forms a gully. A landslide of this type, without any other help than gravity, would be impossible at angles smaller than the angle of repose. However if the $CO_2$ or any other gas is injected into the interior of the granular mass, the grains separate and the gas flow will prevent the contact between the grains, allowing the mixture (sand+gas) to flow like a liquid on any slope (fluidization). The landslide erodes the ground forming a gully and when this is recurrent, whether it is daily or yearly or every time the planet obliquity changes, an evolution of the gullies is expected; this growth in time will dig the deepest channel. The material displaced accumulates at the base of the slopes and forms an apron of debris.

This process is feasible and probable on Mars. In winter and at night, the sand transported by the wind mixes with the $H_2O$ and $CO_2$ snow. In spring or during the day this mixture can warm up, causing the $CO_2$ gas to elevate and flow between the grains of dust, sand and snow, resulting in the fluidization of these materials.

In our model erosion speed of a slope will depends of its material. For instance, on sand dunes gullies will form after few fluidization events. On the other hand in slopes of more hardness material, thousands of events are necessary to form the structures.

### 2.2 Description of the experimental simulations

An experimental device (Fig.1) was built, which consisted of (a) a container with dimensions of 30 cm width, 50 cm length and 30 cm height. Six kg of fine grain volcanic sand (grains between 1.2 mm and 0.03 mm) were placed within the container and used to build a slope (c). To simulate the $CO_2$ on Mars, air at atmospheric pressure was injected through an air pump (d), via a tube (b) of 6 mm diameter and 7 cm length, within the slope. The tube was inserted at a depth of 4 cm, parallel to the slope. So that the air could be injected through the tube it was perforated every 9 mm. When pumping the air the sand bubbles.

To be able to photograph these structures we installed a light source (e) on the left side of the container in such a way that the shadows produced allowed a clear vision of the features created.

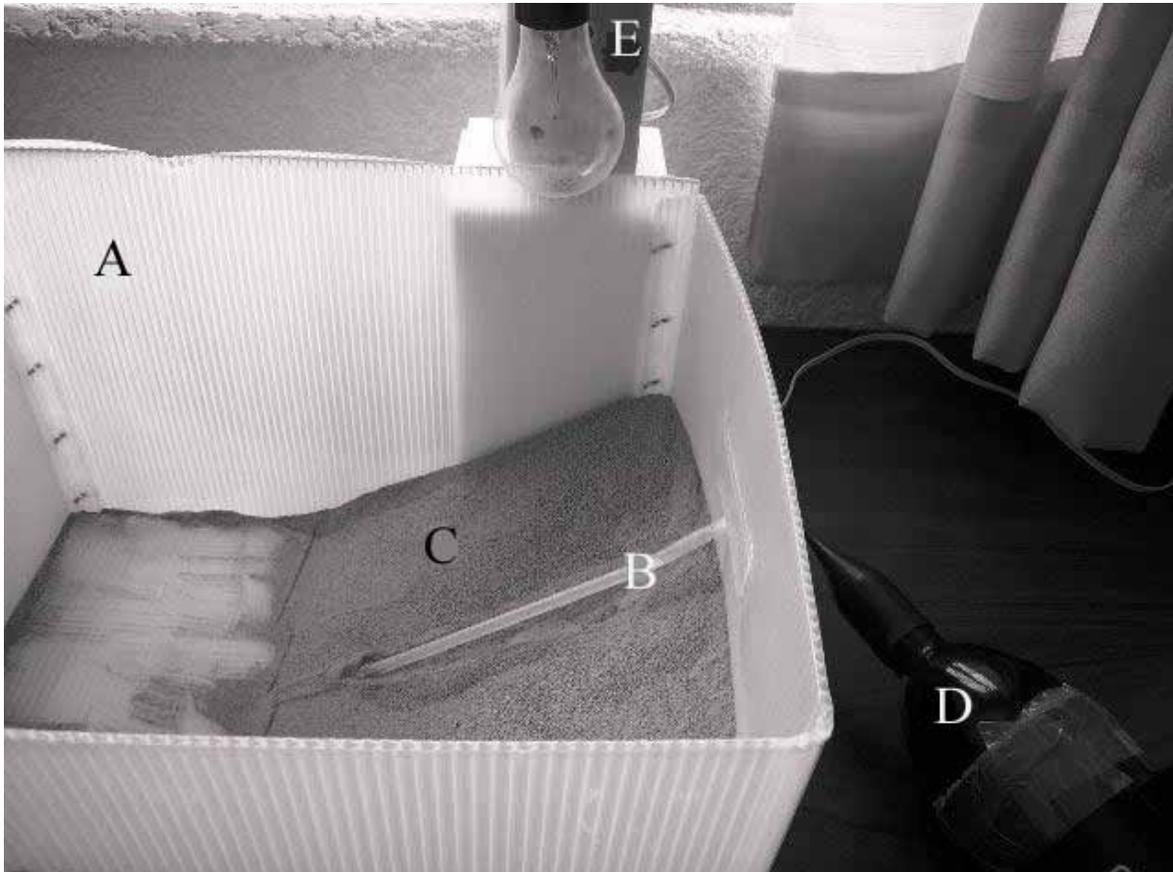

Figure. 1 For the experiments we used this device that consisted of: a) container, b) tube, c) sand slope, d) air bomb, e) light source. The tube perforated within the slope with sand was inserted, air was injected and the sand "flowed" as a liquid.

## 3 Results

The first result of the experimental simulation is that the fluidization allows for the material to slide over the slope, even at angles smaller than the angle of repose.

On injecting the air the fluidization effect of the sand was observed. Also, the structures formed in the laboratory presented similarities to those of the

Martian gullies. In all cases when air was injected, analogue forms of the Martian gullies were created for one event with duration of a few seconds.

In Fig. 2, 3, and 4 some examples of the simulate gullies are compared with some of the Martian gullies; the simulated gully is shown on the left and the real Martian gully on the right. Certain similarities can be appreciated, such as the width of the alcoves and the lobular-shaped aprons.

Fig. 5 shows the result of several fluidization events. The formation of a channel on an apron of debris formed in previous events is observed. Levees formed at sides of the channel as well.

In Fig.2, it can be observed that the alcove and the apron were formed in the simulation gully. The apron has a lobular shape and a narrow channel shorter than the channel of the Martian gully. The Martian gully of this example is located at a mid-latitude of 42.51°N, 312.03°E, and an altitude of -3582 meters over the zero elevation or datum.

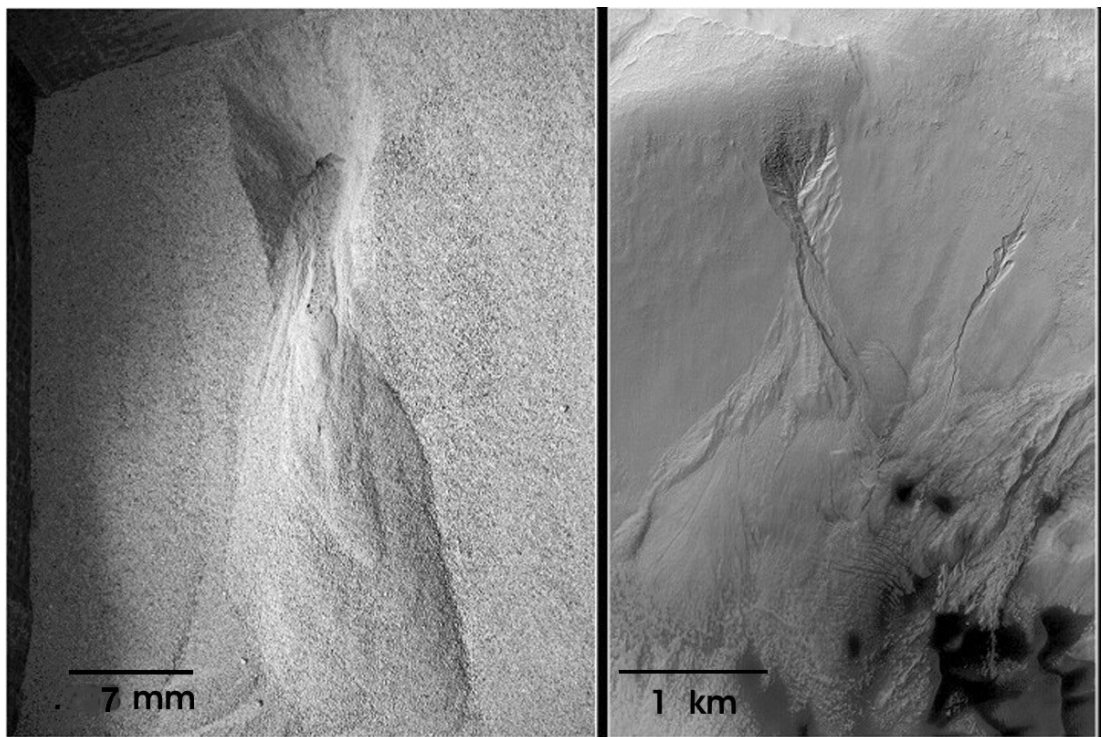

Fig. 2 At right we can see a Martian gully and at left an analog experimental ones. It can be observed that the alcove and the apron formed in the simulation gully are similar to the Martian ones. The apron has a lobular shape and a narrow channel shorter than the channel of the Martian gully. The Martian gully from this example is located at mid-latitude -39.7ºS, 151.4ºE and at altitude of 244 metres. Image MOC- R1500513. NASA/JPL/Malin space Science Systems.

In Fig.3 a large alcove is observed formed by the landslide of material during several stages, as well as some small, narrow channels on the apron. It seems that the Martian gullies also form in stages. In this example, the gully is located at a latitude of -39.7°S, a longitude of 151.4°E and at an altitude of 244 meters.

In Fig. 4 we show a gully with a large alcove and branching structures, in the experimental gully, as are formed in the case of Martian ones. This gully is located at a high latitude of 70.9°S, 20.7°E and at an altitude of 878 meters.

To observe if in the model the gullies are formed with an inclination smaller than the angle of repose some additional experimental simulations were performed, varying the angle of the slope between 10° and 35°. The slope angle was changed measuring the θ angle every time and photographing every variation. Subsequently, we measured, on the photograph, the length of the apron from the end of the bubbling zone until the lowest part of the apron. When a stable state for the simulation was obtained, defined by the moment that the material did not slide anymore, a horizontal area within the gully alcove formed, where the injected air continued bubbling without provoking anymore landslides.

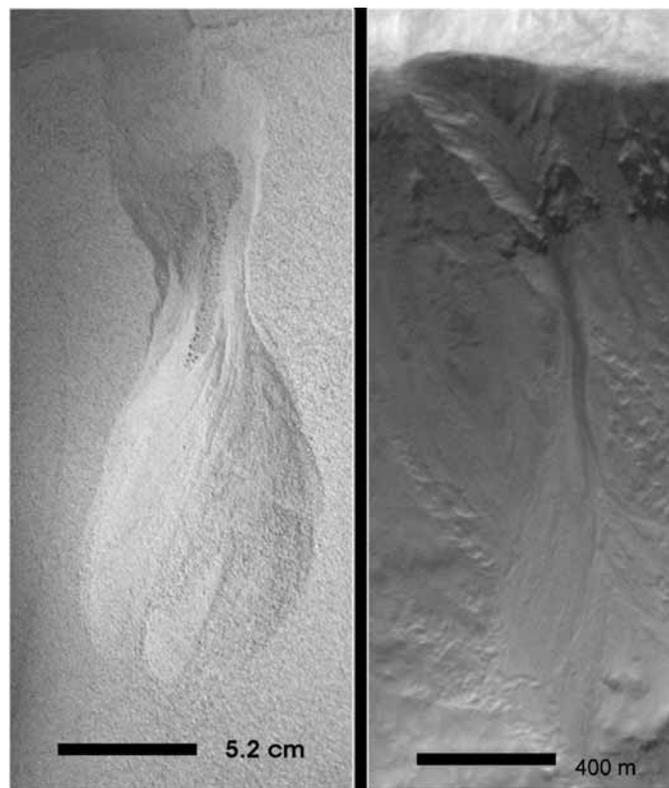

Fig. 3 Another Martian gully (right) and an analog experimental gully (left). In the experiment a large alcove is observed formed by the landslide of material during several stages, as well as some small, narrow channels on the apron. The Martian gully is found at mid-latitude 42.51ºN, 312.03ºE, and at altitude of -3.582 metres below zero elevation or datum. Hirise image PSP_001 389_2225. NASA/JPL/UNIVERSITY OF ARIZONA.

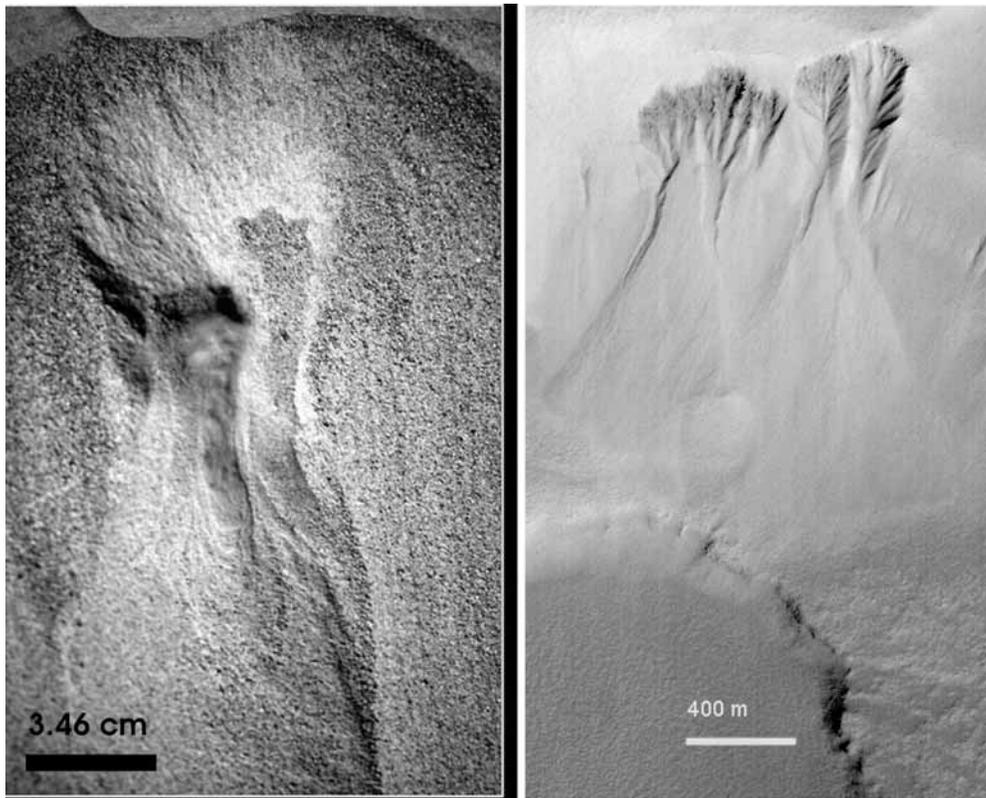

Fig. 4 Another Martian gully (right) and an analog experimental gully (left). Both gullies have a large alcove and branching structures. This Martian gully is located at a high latitude of -70.9ºS, 20.7ºE and at altitude of 878 metres. Image MOC- E1000770. NASA/JPL/Malin space Science Systems.

This area was denominated as the bubbling zone. The length of the apron (L) was normalized to the length of the bubbling zone ($L_0 \sim 8$ cm). The results can be observed in table I and in Fig.5. The normalized length according to the angle of inclination is defined by the following equation:

$$\frac{L}{L_0} = 0.0038\theta^2 - 0.0134\theta \qquad (1)$$

Where $\theta$ represents the inclination in degrees.

The results of this experimental simulation indicate that the length of the debris apron is a quadratic function of the angle of inclination and that fluidization causes the material to slide, even when the angle of inclination is less than the angle of repose.

Table I

The length of the apron (L) normalized throughout the bubbling zone ($L_0$)

| Probe | Angle | $L/L_0$ |
|---|---|---|
| 1 | 0 | 0 |
| 2 | 10 | 0.42 |
| 3 | 16 | 0.65 |
| 4 | 21 | 1.33 |
| 5 | 25 | 1.62 |
| 6 | 27 | 2.73 |
| 7 | 33 | 3.51 |

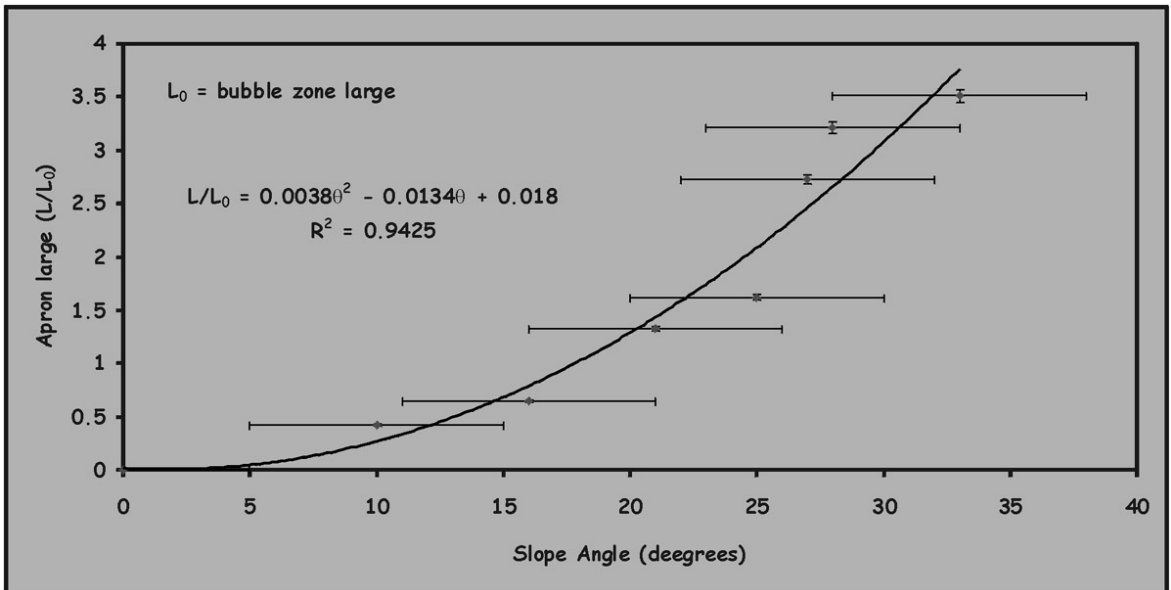

Figure 5 Apron length vs. slope angle. The length of apron (L) is normalized to the length of the zone of bubbling ($L_0$).

4 Discussion

A particular process such as acoustic fluidization has been proposed by Melosh (1979) to explain the generation of dry flow over the surface of other planets. On Earth acoustic fluidization generates the falling of material in small angles due to seismic activity, but on Mars this process takes place due to certain tremors or due to shock waves caused by the impact of meteorites. The acoustic fluidization process would be capable of producing gullies only in non consolidated material since these events (the acoustics) would be unique events, hence not capable of forming gullies in areas that require recurrent events. However it cannot be considered that many gullies are formed in association with this phenomena since the frequency of meteorite impacts has been reduced during geologic history of the planet and there seems to be no evidence of seismicity on Mars.

Due to the fact that it is not possible to simulate all the Martian environmental conditions in the laboratory, the experimental simulations only focused on reproducing analogue structures to the Martian gullies.

Our model suggests that Martian gullies can form without the presence of water or any other liquid. The results of experimental simulations indicate that fluidization is an adequate mechanism to generate similar features to Martian gullies.

With this mechanism, no other special conditions are required as in the case of the models that invoke the water. The materials and conditions necessary are common on the surface of Mars: dust, $CO_2$ and temperature changes extreme enough to allow for $CO_2$ to change from gas to solid and then again solid to gas.

In the experimental simulations structures similar to a gully are formed in only one fluidization event due to the fact that the fluidized material is the same as the one that forms slopes. On Mars this can happen in sand dunes. But on solid rock, like the slopes of the Martian impact craters, the gullies must form with multiple fluidization events during variable periods of time. In these places the dust deposited by the wind erodes the slope and not the slope's own material, as in the case of the simulations.

In the simulation the structures formed by just one flow can be considered as examples of what would happen on Mars, after a series of fluidization events during several daily or seasonal cycles.

Our simulations cannot be seen as avalanches or landslides since, as well as the force of gravity, gas also intervenes diminishing the friction between the grains of dust.

To compare the similarities between the flow of the experimental simulations and the real flow in the Martian environment the Reynolds number was calculated in both cases. The Reynolds number is defined as (Fox y McDonald, 1985)

$$R_e = \frac{DV\rho}{\mu} \qquad (2)$$

In this case D is the width of the channel; V is the velocity of the flow, $\rho$ the density and $\mu$ the viscosity.

Mangold et al (2003) provide the following relation for viscosity

$$\mu = \frac{(\rho g \sin\alpha)T^2}{2V} \quad (3)$$

Where g is the gravity acceleration, $\alpha$ the angle of inclination and T the channel depth that we considered as T=0.5 D.

Using equations (2) and (3) and obtaining V for the slide on an inclined plane we get

$$R_e = \frac{16S}{D} \quad (4)$$

Where S is the length of the channel, this equation is important because it allows us to evaluate Reynolds number by only measuring the length and width of the channel in a photograph.

For the simulation Re was between 130 and 224. In the Martian gullies all the values for the Reynolds number evaluated with the equation (4) were of the same order of magnitude (see table II). Of the 20 Martian gullies for which this evaluation was performed 12 remained within the range of values of the simulation, confirming that both, experimental and Martian, flow systems are similar. Coleman et al. (2009) obtained in their experiments with water the value for the Reynolds number ~2000 which is very high for the values calculated for the Martian gullies, demonstrating that water does not form these.

One condition which must fulfilled by every model is the refilling of the system after every flow event to allow for repetitive events which can form gullies. This is still an unsolved problem for the models based on water since a mechanism refill does not exist for aquifers on Mars (Kargel, 2004). In our model this problem is solved automatically since the refilling of sand and dry ice comes from the atmosphere and the energy necessary for the process comes from solar heating.

Table II
Reynolds number for some Martian gullies
A = Gully and Image, La and Lo = Latitude and longitude on Mars, S Length of the channel, D = Wide of the Channel, $R_e$ = Reynolds number
(Length and wide are measured over the images then it is not the real distances in gullies)

| A[1] | La, Lo | S(cm) | D (cm) | $R_e$ |
|---|---|---|---|---|
| [2]Gully1 | La 36°, Lo 351.9° | 4 | 1 | 64 |
| [2]Gully2 | La 36°, Lo 351.9° | 4.5 | 0.3 | 240 |
| [2]Gully3 | La 36°, Lo 351.9° | 2.1 | 1.7 | 19 |
| [3]Gully1 | La-40.4°,Lo 155.3° | 6.2 | 0.5 | 198 |
| [4]Gully1 | La 36°, Lo 351.9° | 3.2 | 0.3 | 171 |
| [4]Gully2 | La 36°, Lo 351.9° | 4 | 0.3 | 213 |
| [5]Gully1 | La -36°, Lo 199.5° | 5.8 | 0.9 | 103 |
| [5]Gully2 | La -36°, Lo 199.5° | 3.6 | 0.7 | 82 |
| [5]Gully3 | La-36.0°,Lo 199.5° | 3.6 | 0.7 | 82 |
| [5]Gully4 | La-36.0°, Lo199.5° | 4.3 | 0.5 | 138 |
| [6]Gully1 | La-38.9°, Lo195.9° | 5.4 | 0.6 | 144 |
| [7]Gully1 | La -68.5°, Lo 1.3° | 5.4 | 0.2 | 432 |
| [7]Gully2 | La -68.5°, Lo 1.3° | 8.1 | 0.2 | 648 |
| [7]Gully3 | La -68.5°, Lo 1.3° | 7.0 | 0.3 | 373 |
| [8]Gully1 | La-38.8°, Lo182.1° | 3.5 | 0.1 | 56 |
| [8]Gully2 | La-38.8°, Lo182.1° | 4.0 | 0.2 | 320 |
| [9]Gully1 | La -68.5°, Lo 1.3° | 5.3 | 0.1 | 85 |
| [9]Gully2 | La -68.5°, Lo 1.3° | 3.6 | 0.3 | 192 |
| [9]Gully3 | La -68.5°, Lo 1.3° | 3.6 | 0.3 | 192 |
| [9]Gully4 | La -68.5°, Lo 1.3° | 4.1 | 0.2 | 328 |

[1] http://www.uahirise.org/katalogos.php
[2] Image PSP-010354-2165, [3] MOC 1609, [4] ESP_014127_1420, [5] ESP_14263_1435
[6] ESP-013894-1410-RED.NOMAP, [7] ESP_013097_1115_RED.NOMAP
[8] ESP_013288_1410_RED.NOMAP, [9] ESP_013097_1115_RED.NOMAP,

The reason why most of the gullies are located in mid-latitudes can also be explained with the model. With the conditions on Mars (temperatures between 133K and 293K; pressure between 3 and 12.5 mb), the $CO_2$ sublimates between the 140 K and 160 K. In equatorial zones (Balme et al. 2006), temperatures are high enough so that a great part of the time $CO_2$ snow is not produced and so the fluidization events are scarce. In polar zones temperatures are very low, and sublimation of $CO_2$ is much slower and the fluidization events will also not be very frequent. At mid-latitudes, the night winter temperatures are low enough to produce $CO_2$ snow and the daily summer temperatures are high enough so that the solid $CO_2$ sublimates quickly. Hence the mid-latitudes have optimal conditions for the fluidization events to be more frequent.

In the experimental simulations it can be stated that the alcove and debris apron form, but it is not always possible to observe the channel. Actually the channel forms most of the times but gets blocked by the fluidized sand. In the Figure 6A the flow going through a channel (arrow zone) which further down opens into an apron can clearly be seen. Fig. 6B shows the same experiment but fractions of second later. Part of the alcove crumbles producing a much larger flow of sand and the channel starts to get blocked. Fractions of second later (Fig. 6C) the flow is less and the channel can hardly be seen. In the fig. 6D the flow has ended but the channel is completely obstructed.

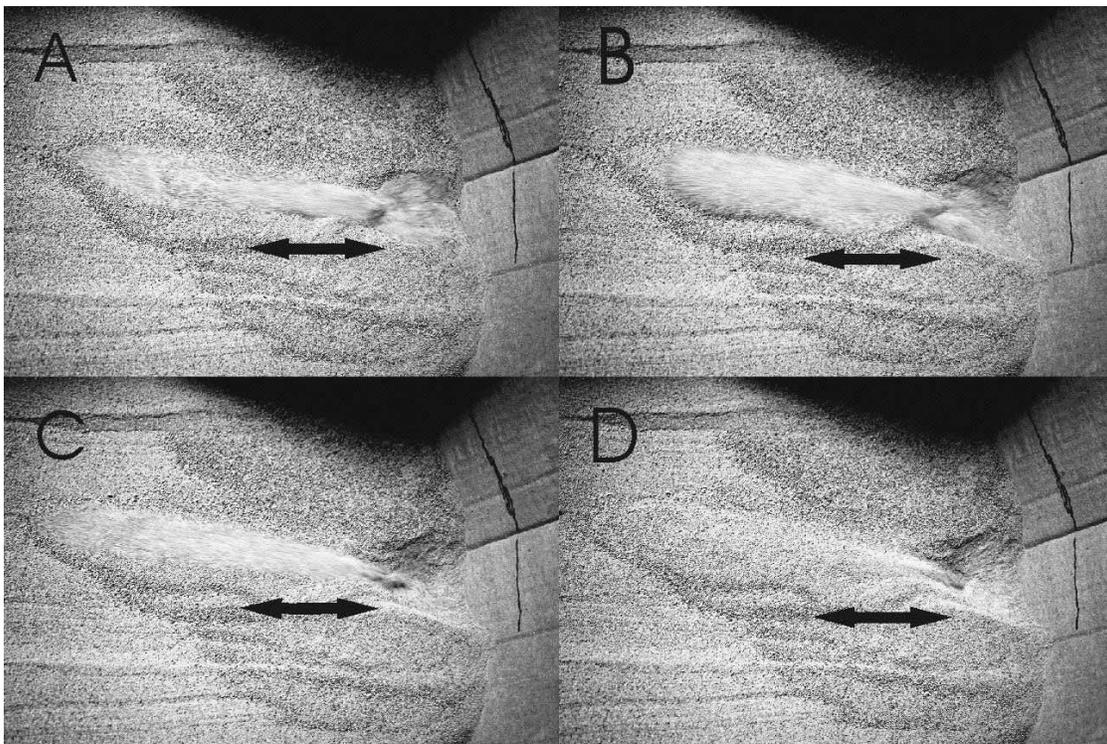

Fig. 6 In the Figure 6A the flow going through a channel (arrow zone) which further down opens into an apron can clearly be seen. Fig. 6B shows the same experiment but fractions of second later. Part of the alcove crumbles producing a much larger flow of sand and the channel starts to get blocked. Fractions of second later (Fig. 6C) the flow is less and the channel can hardly be seen. In the fig. 6D the flow has ended but the channel is completely obstructed.

The fluidization events on Mars exhibit an important difference in comparison with laboratory analogues: in the experimental simulations air is injected into the bubbling zone and fluidization lasts until the air comes out of the sand. In contrast, the fluidized material on Mars also transports solid $CO_2$ which, with friction caused by the sliding action, sublimates on its journey making the fluidity time longer, and forming long channels like the ones observed in some Martian gullies. It is for this reason that the channels that were formed in the simulation are short when compared with those on Mars, since, in experimental conditions, it is impossible to refill the sand with gas during its advance.

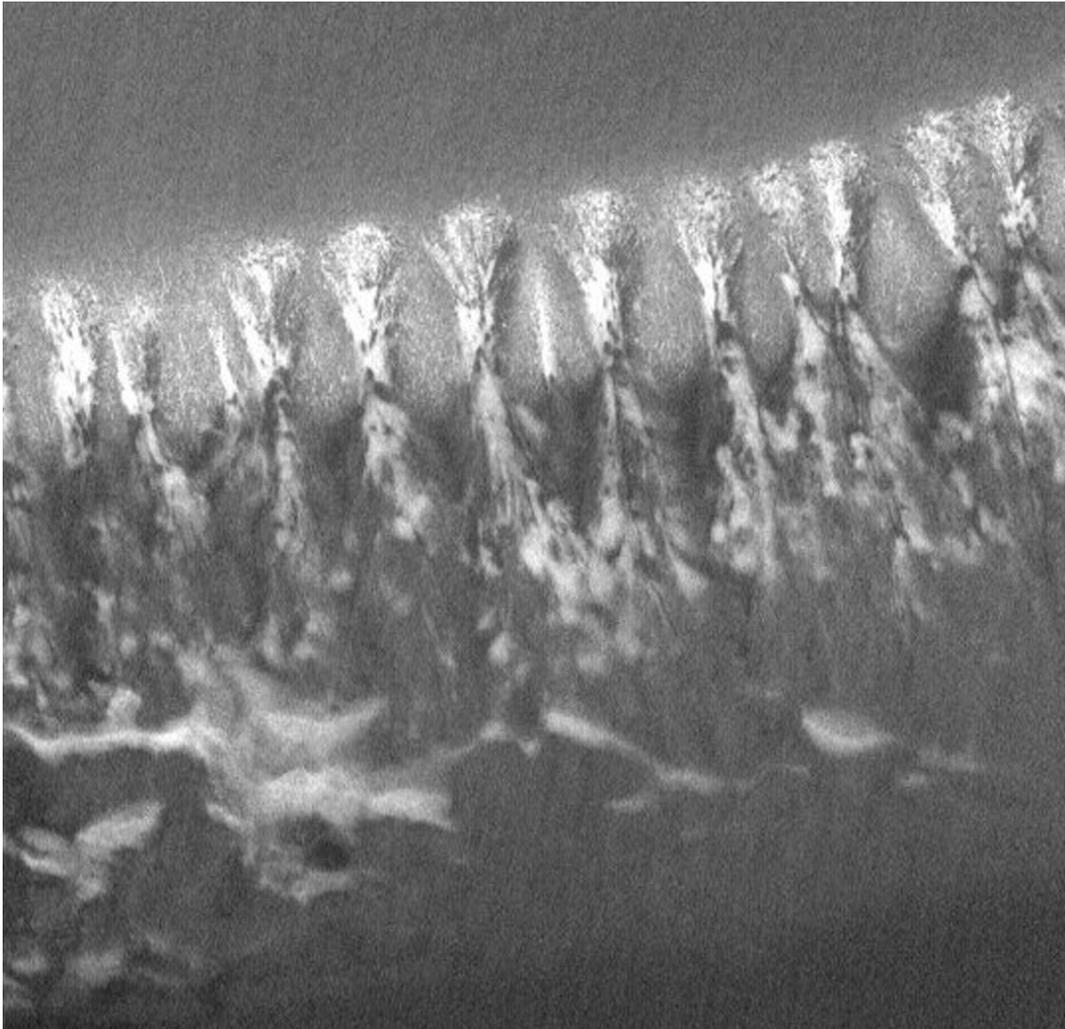

Fig. 7 Some gullies in a Martian slope with $CO_2$ snow deposits. This would be the scenario prior to a fluidization. M0906352 (Hartmann, 2003).

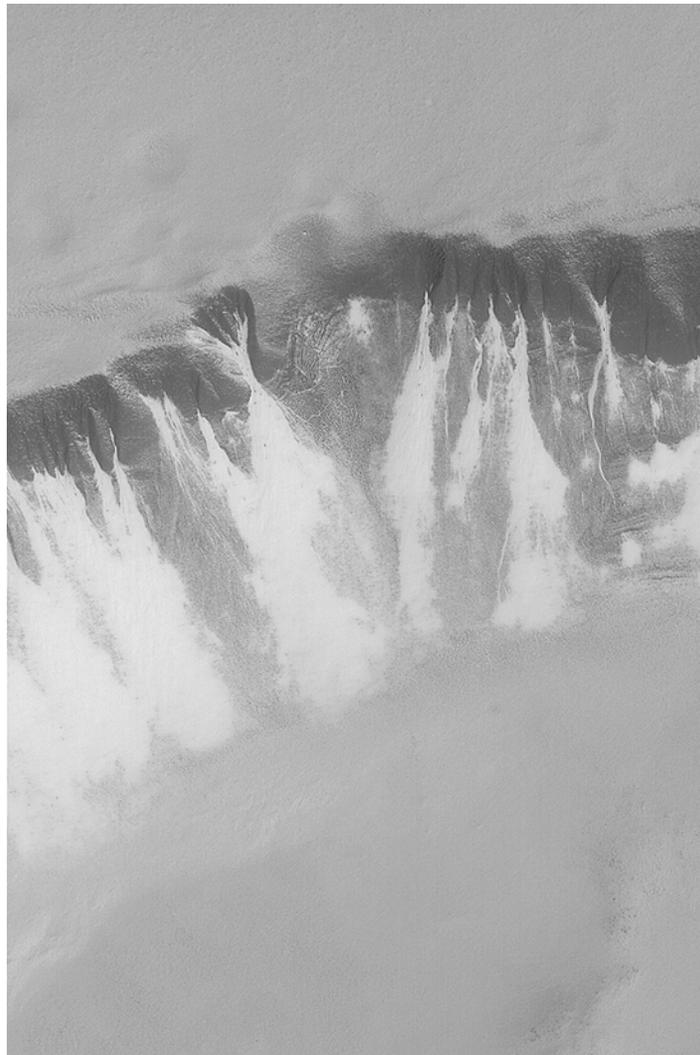

Fig. 8 The alcoves are empty of frozen material and in the slope there are fans of white material. This scenario is congruent with that which would result following a fluidization event. MOC2-467 $CO_2$ snow deposit in a slope with gullies. 70.7°S, 358.2°W. Credits. NASA/JPL/MSSS.

This phenomenon also explains why in the simulations the channel is obstructed and on Mars it is not. On Earth certain landforms exist, similar in shape but not in size to the Martian gullies and are produced by the fluidization phenomenon. For example, some gullies on the volcano slopes are formed by pyroclastic flows. These flows consist of volcanic ashes fluidized by hot volcanic gases (Tarbuck and Lutgens, 1999). Another landforms formed by fluidization consists of the submarine gullies which form on the continental slope (Tarbuck and Lutgens, 1999; Salles et al, 2008). These landforms are formed by turbidity currents which basically consist of marine sediment fluidized by water. The appearance of these submarine gullies is very similar to the Martian gullies, even though they are much longer, probably caused by the water being much more dense than the $CO_2$ gas, which is capable of keeping the particles elevated for a longer period of time. Due to this the fluidized material is capable of flowing a greater distance and forming very long channels.

Heldmann and Mellon (2004) reported that most of the Martian gullies are on slopes with an inclination smaller than the angle of repose. One of the most important things about our model is that it allows the formation of gullies with angles with these inclinations. In Fig. 5 it was observed that all the angles measured were smaller to than angle of repose (~ 34°), demonstrating that fluidization is capable of generating similar features to gullies at small angles of inclination.

Stewart and Nimmo (2002) studied the possibility of sustained flow by gas from subterranean liquid or solid $CO_2$ deposits. They concluded that a typical apron consists of ~$10^4$ $m^3$ of material (Malin and Edgett, 2000). For fluidize this material it is necessary ~ $10^7$ $m^3$ of gaseous $CO_2$, the equivalent to ~$10^5$ kg of solid $CO_2$. These numbers are consistent with the proposed model, where an alcove with a surface of $10^4$ $m^2$, containing a dust deposit, $CO_2$ and $H_2O$ snow with a depth of one meter, will form $10^4$ $m^3$ of this mixture. If 0.6% of this mixture was $CO_2$ snow then we would have ~ $10^5$ kg of $CO_2$, which would sublimate and fluidize the rest of the material, would deposit ~$10^4$ $m^3$ in the debris apron. Of course this deposit does not necessarily have to be produced by only one fluidization event. The study performed by Stewart and Nimmo (2000) is based on the underground $CO_2$, deposits. However considering the previous numbers it can be deduced that $CO_2$ deposited from the atmosphere would be enough.

During the periods of high inclination on the Mars axis, it can be expected that great quantities of $CO_2$ ice will be deposited during the long winter nights (~ 1 year), which will sublimate over the summer. As a result of this over the years it is hoped that the fluidization mechanism will be more efficient.

Fig. 7, shows that the alcove has accumulated frozen material. Although it is not known if this consists of $H_2O$ or $CO_2$ ice, or a mixture of both, the situation is adequate to originate a fluidized landslide, if there is dry ice in the mixture. This would represent the scenario previous to a fluidization event. On the other hand in Fig. 8 the alcove lacks the frozen material and at the bottom of the slope some clear color aprons can be seen. Although in this case we nor known if it is $H_2O$ and $CO_2$ snow or a mixture of both, the scenario is consistent with that which results after the fluidization event.

Recently Malin et al., (2006) described an event produced on a slope which was interpreted as a water flow which left its sliding trail. A whitish substance seems to have left this trail; this could be water or according to our scenario, it could be $CO_2$ or $H_2O$ snow and sand sliding in an event of fluidization. The existence of the gullies without the intervention of liquid water is possible, since recently the existence of gullies on the moon (Bart, 2006), where it is evident that liquid water does not exist on the surface, due to the extreme conditions of the satellite. The fact that these features are similar to those on Earth, the Moon or Mars is no proof that the same processes have intervened the origin and formation.

In conclusion: the proposed model contemplates the formation of similar structures to Martian gullies.

The model explains better the formation of Martian gullies at different latitudes, on different geological terrain, and with different orientations, since the formation processes are external (material from the atmosphere) and according to the actual conditions on the Mars planet.

Our model does not require exotic conditions; only the ones found at present on Mars. The refilling of the system to produce a renovated flow is an implicit process in the model and finally this mechanism allows for the gullies to form on the slopes at angles smaller than the angle of repose.

Acknowledgements.

The analogue gullies images to the experimental simulation were provided by The Center for Earth and Planetary Studies (CEPS), National and Space Museum, Smithsonian Institution.